\documentclass[conference]{IEEEtran}
\IEEEoverridecommandlockouts

\usepackage{cite}
\usepackage{url}
\usepackage{bbm}
\usepackage{amsmath,amsfonts} 
\usepackage{float}
\usepackage{graphicx}
\usepackage{algorithm}
\usepackage{algpseudocode}
\usepackage{siunitx}
\usepackage[labelfont=bf]{caption}
\usepackage{subcaption}
\usepackage{tikz}
\usepackage{pgfplots}
\usepackage{epstopdf}
\usepackage{multirow}
\usepackage[font=scriptsize]{caption}
\usepackage{soul}
\usepackage{color, colortbl}
\usepackage{array}
\usepackage[T1]{fontenc}
\usepackage[utf8]{inputenc}
\usepackage{mathptmx}
\usepackage{listings}
\usepackage{adjustbox}
\usepackage{makecell}

\usepackage{enumitem}
\usepackage{hyperref}
\usepackage{pifont}

\def\BibTeX{{\rm B\kern-.05em{\sc i\kern-.025em b}\kern-.08em
    T\kern-.1667em\lower.7ex\hbox{E}\kern-.125emX}}

\begin{document}

\title{\textit{ExpIDS}: A Drift-adaptable Network Intrusion Detection System With Improved Explainability}


\author{\IEEEauthorblockN{Ayush Kumar, Kar Wai Fok and Vrizlynn L.L. Thing}
\IEEEauthorblockA{Cyber Security Strategic Technology Centre \\
ST Engineering \\
Singapore, Singapore \\
Email: ayush.kumar@u.nus.edu, fok.karwai@stengg.com, vriz@ieee.org}
\thanks{Corresponding author: A. Kumar (email: ayush.kumar@u.nus.edu).}}
			


\maketitle

\begin{abstract}
Despite all the advantages associated with Network Intrusion Detection Systems (NIDSs) that utilize machine learning (ML) models, there is a significant reluctance among cyber security experts to implement these models in real-world production settings. This is primarily because of their opaque nature, meaning it is unclear how and why the models make their decisions. In this work, we design a deep learning-based NIDS, \textit{ExpIDS} to have high decision tree explanation fidelity, i.e., the predictions of decision tree explanation corresponding to ExpIDS should be as close to ExpIDS's predictions as possible. \textit{ExpIDS} can also adapt to changes in network traffic distribution (drift). With the help of extensive experiments, we verify that \textit{ExpIDS} achieves higher decision tree explanation fidelity and a malicious traffic detection performance comparable to state-of-the-art NIDSs for common attacks with varying levels of real-world drift.

\end{abstract}


\section{Introduction}
\label{intro}

Ensuring the security of networks against intrusions is crucial, and network intrusion detection systems (NIDS) are key to achieving this goal. Historically, NIDSs have primarily utilized signature-based techniques \cite{bro, snort}, which involve matching attack patterns to pre-existing signatures. These methods, however, necessitate considerable manual effort from experts to develop and update these signatures. Additionally, they struggle to detect new, previously unknown attacks. The advent of machine learning (ML)-based NIDSs, which use ML models to strengthen existing security mechanisms, has resulted in remarkable performance and accuracy \cite{dd-nids, MLIDissues}. Unlike traditional, signature-based approaches, ML-based NIDSs do not require manual creation of signatures beforehand. Unsupervised ML-based NIDSs can even alert to previously unrecognized attacks. Owing to these benefits and the rapid advancement of ML technologies, ML-based NIDSs have made significant progress in recent times.

Though ML-based NIDSs show great potential, network security practitioners are often hesitant to use them in real-world settings \cite{dos-n-donts-ml}. The main issue is the opaque nature of these solutions, making it hard to understand the reasons behind their decisions, unlike simpler rule-based methods that security professionals are more familiar with. Without unambiguous explanations of the underlying ML models' decisions, building security practitioners' trust in ML-based NIDSs is challenging. The ML models' opaque nature can allow hidden correlations from unrelated data to go unnoticed, leading to a false sense of the NIDS’s effectiveness. Utilizing explanation tools can reveal these hidden correlations, enabling security practitioners to evaluate their effect on the performance of the NIDS.

In our previous work on evaluating the explainability of state-of-the-art deep learning-based NIDSs \cite{ayush-iot-nids-explain}, we had observed that some NIDS models can have better interpretability in terms of decision trees (higher fidelity to original black-box model) than other NIDS models. One of the reasons to which we attributed this observation was the complexity of the underlying NIDS model's architecture. Therefore, in this work, we carefully design a deep learning-based NIDS, \textit{ExpIDS} to have high decision tree explanation fidelity which means that the predictions of decision tree explanation corresponding to ExpIDS should be as close to ExpIDS's predictions as possible. ExpIDS uses a single de-noising autoencoder and can adapt to changes in network traffic distribution (drift). 
Through extensive experiments, we verify that ExpIDS achieves higher decision tree explanation fidelity while maintaining a malicious traffic detection performance comparable to state-of-the-art NIDSs for common attacks with varying levels of real-world drift.

The main contributions of our work are as follows:
\begin{itemize}
	\item We design a deep learning-based NIDS, \textit{ExpIDS} which lends itself better to explanation.
	\item We verify that \textit{ExpIDS} has higher decision tree explanation fidelity compared to state-of-the-art NIDSs.
	\item We evaluate the malicious traffic detection performance of \textit{ExpIDS} and compare it with state-of-the-art NIDSs for common attacks with varying levels of real-world drift.
\end{itemize}
 

\section{Background}
\label{background}
In this paper, since we design our proposed NIDS using an autoencoder (AE), it is essential to briefly explore the AE's structure and operation. Further, since our proposed NIDS is designed for better explainability, we present an overview of the xAI tool used for generating explanations.
\subsection{Autoencoders (AEs).} 
\begin{figure}[h]
\centering
\includegraphics[scale=0.35]{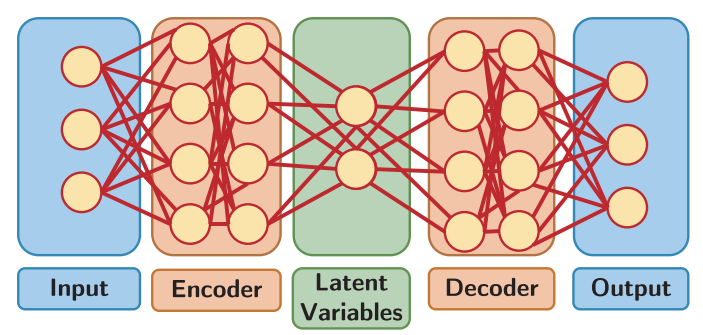}
\caption{Autoencoder Architecture}
\label{autoenc}
\end{figure}

Autoencoders \cite{ae-paper}, rooted in artificial neural networks, serve as fundamental classification algorithms for NIDSs \cite{kitsune, cade} (see Fig. \ref{autoenc} for their architecture). Their input and output layers share the same number of nodes. The core idea behind AEs is to iteratively align input and output ($\vec{x}$ and $\vec{y}'$ respectively) during training, preserving sample characteristics within the AE’s neurons. Internally, they feature a hidden layer $h$ (can be more than one hidden layers), a code representing the input. Conceptually, an AE comprises two parts: an \textit{encoder} function $h = f(x)$ and a \textit{decoder} responsible for reconstruction $r = g(h)$. While AE resemble a specialized form of feedforward networks, their training involves techniques like mini-batch gradient descent and back-propagation.

\subsection{Blackbox Explanation Techniques}
\label{background-xai}
The end users of deep learning-based NIDS solutions offered by cybersecurity service providers typically do not have access to the underlying models and their parameters in order to keep NIDS implementation details private. Henceforth, it is assumed that the NIDSs discussed in this work are available only as blackbox systems, and xAI techniques are employed to explain the decision-making behavior of such systems. We will now provide a brief overview of TRUSTEE, the blackbox xAI tool that we used in this study. TRUSTEE \cite{trustee} provides global-level explanation for the blackbox model's decisions. We do not use tools such as SHAP/LIME \cite{shap, lime} which provide local-level explanation by analyzing only a subset of the blackbox model’s decisions as those explanations might be misleading. Here, local-level explanations clarify why the ML model underlying a particular NIDS makes certain decisions for individual inputs. On the other hand, global-level explanations identify key features derived from the ML model and describe the connections between feature values and the overall classification process. 

\textbf{Overview of TRUSTEE}: TRUSTEE is a state-of-the-art global-level explanation tool designed for ML-based network security solutions.  It takes the ML model as an input and outputs a corresponding interpretable decision tree (DT). This tool can identify different types of inductive biases in blackbox ML models such as sensitivity to out-of-distribution data samples and spurious correlations. The key points of the DT explanation created by TRUSTEE are compiled into a trust report, aiding end users in assessing underspecification issues with the blackbox model and hence, its reliability.

In the TRUSTEE algorithm, there is an inner loop which is designed to generate different high fidelity DT explanations, one per iteration. It does so by applying a teacher-student dynamic derived from imitation learning that uses the blackbox model, $\pi^*$ as an oracle in conjunction with a carefully curated dataset $D’$ (usually a fraction of the training dataset) to guide the training of a surrogate ``white-box'' model in the form of a DT that imitates the black-box’s decisions. In contrast, the purpose of the outer loop is (i) to select from among the $N$ high fidelity DTs that have been generated in the process of executing the inner loop the DT with the highest fidelity, (ii) to transform this resulting DT into a high-fidelity and low-complexity DT by means of a post-processing step that consists of applying a purposefully developed pruning method called \textit{top-$k$ pruning} which carefully scrutinizes only the top $k$ branches of an extracted high-fidelity DT, ranked by the number of input samples a branch classifies, and (iii) to consider all $S$ high-fidelity and low-complexity DTs that have been generated in the process of executing the outer loop and output the one that is the most stable in the sense of having the highest mean agreement among these $S$ DTs. Here, \textit{fidelity} is defined as the R-squared value between blackbox model’s predictions and those obtained by the DT explanation.

\section{\textit{ExpIDS} Architecture and Operation}
\begin{figure*}[t]
	\centering
	\includegraphics[scale=0.4]{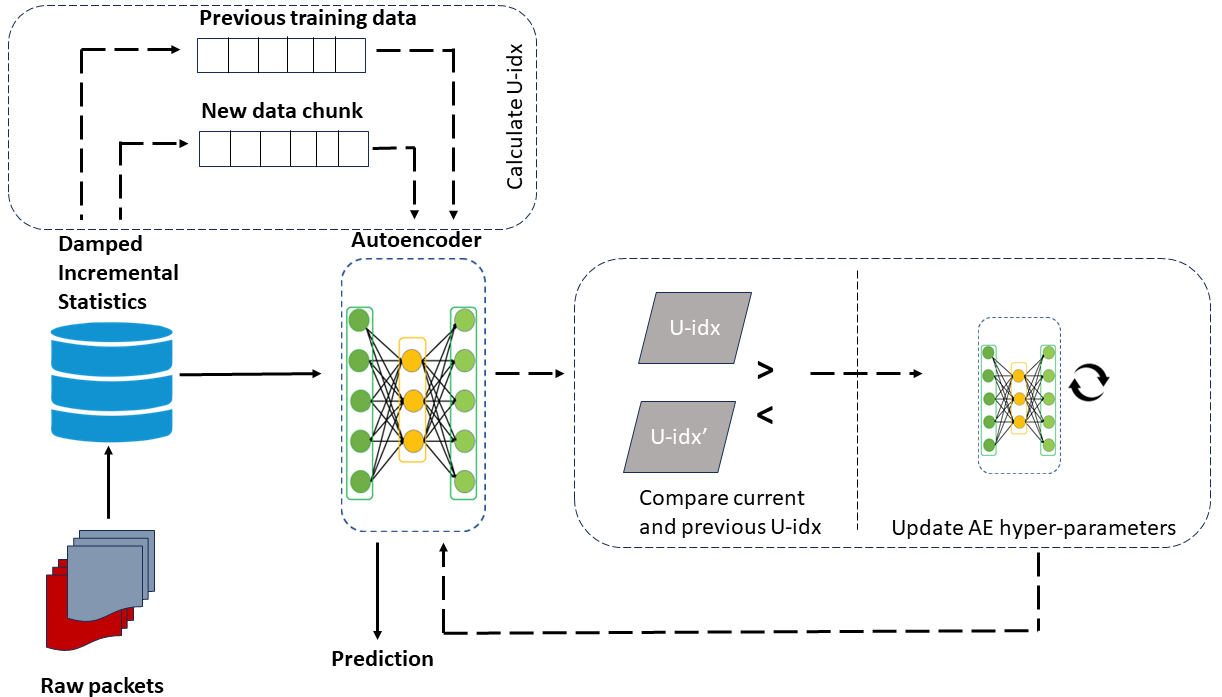}
    \caption{ExpIDS Architecture. Damped incremental statistics method is used to extract features from raw packets captured and then a single de-noising autoencoder (AE) predicts whether the packets are normal or anomalous. The training data used for previous AE update and newly arrived data (after release of ground-truth labels) are used to verify any changes in network traffic distribution and accordingly update the AE hyper-parameters.}
    \label{expids-arch}
\end{figure*}

In our previous work \cite{ayush-iot-nids-explain}, based on the TRUSTEE DT fidelity scores obtained for state-of-the-art deep learning-based NIDSs, we had recommended that to build an NIDS with better ``explainability'', we should use less complex ML architectures and conventional components. Based on that recommendation and other considerations, we design our proposed NIDS, \textit{ExpIDS} whose architecture is shown in Fig. \ref{expids-arch}. 

\subsection{Feature Extraction}
We use the features extraction framework with damped incremental statistics (DIS) proposed in \cite{kitsune}. DIS consumes less storage and forgets the contribution of older traffic instances to the incremental statistics over time. It does so by calculating incremental statistics over a damped temporal window which uses an exponential decay factor. When the dampening weight for a traffic instance's feature value corresponding to an incremental statistic reaches zero, the value is deleted from memory. This feature extraction technique is suitable for high-speed packet traffic in real-world networks. The feature extraction framework extracts both 1-D and 2-D packet-level statistics over multiple temporal windows, where 1-D statistics capture a packet transmitter's traffic behavior during a connection while 2-D statistics capture the relationship between the two-way traffic between a packet transmitter and receiver during a connection.  

An alternative to DIS is the \textit{ip2V} vector embedding technique from ENIDrift NIDS \cite{enidrift}. ip2V is an incremental feature extraction method based on a three-layer neural network which captures both apparent and hidden releationships between related packets. However, we do not use it for \textit{ExpIDS} since the ip2V embedding vectors included in the explanation generated for an NIDS can not be interpreted in terms of meaningful traffic features for security practitioners.

\subsection{Packet Traffic Anomaly Detection}
We use a single unmodified de-noising autoencoder (AE) which takes the input (feature vector extracted from incoming network packet) from the feature extraction module and makes a decision on whether the packet is normal or anomalous. Our design choice for the main traffic classification module stems from the observation that modified AEs and AEs used in combination with other neural networks (NNs) are shown to have low decision tree (DT) explanation fidelity to original black-box models in our previous work \cite{ayush-iot-nids-explain}.

During the training phase, the AE is trained on normal traffic features where it learns to reconstruct those features while minimizing the reconstruction error. During the execution phase, the AE detects malicious traffic by comparing the corresponding reconstruction error with a threshold (maximum of all reconstruction errors calculated for the training dataset). An AE trained on normal traffic features produces a large error when it encounters malicious traffic.

\subsection{Network Drift Detection and AE Update}
As explained in \cite{enidrift}, the real-world network traffic is subject to drift (change in distribution of underlying packet statistics) due to factors such as imbalanced traffic, ML attacks and concept drift. Imbalanced traffic refers to the issue that in real-world network traffic, the number of normal and anomalous packets are not equal and in fact, the ratio between them might change over time. ML attacks include adversarial attacks which use well-designed malicious data samples to cause ML-based IDSs to mis-classify those samples as benign and data contamination attacks which contaminate IDS training datasets with falsely labelled data to degrade the IDS's performance. Concept drift refers to the change in normal network activities or the introduction of new attack types which might not be present in the original IDS training dataset.

\begin{algorithm}[h]
	\centering	
	\caption{\scriptsize{Calculate U-idx}}
	\label{A1}
	\begin{algorithmic}[1]
		\State \textbf{INPUT}: $D\_c$, $D\_t$, $K$, $K'$, $M$
		\State \textbf{OUTPUT}: $U$
		\If {$D\_t$ is renewed}
		    \For {$k=1$ to $K$}
		         \State Take $M$ samples from $D_t$ $\rightarrow$ $D'$
		         \State Train prev. AE on $D'$ $\rightarrow$ $r_k$
		    \EndFor
		\EndIf
		\For {$d$ in $D_c$}
		    \For {$k'=1$ to $K'$} 
		        \State $o_k^{'}$ $=$ thresh\_comp($r_{k'}(d)$)
		        \State $e_k^{'}$ $=$ err($o_k^{'}$)   
		    \EndFor
		    \State $U_i$ $=$ $w.Var(o') + (1-w).Mean(e')$ 
		\EndFor
	\end{algorithmic}
\end{algorithm} 

To address network drift, we use an update index (\textit{U-idx}) whose value is calculated using the Algorithm \ref{A1}. The U-idx evaluates incoming data chunks (newly arrived data combined with previous data) based on its stability (variance of predictions) and accuracy (error of predictions) calculated by training AE with previous hyper-parameters (see line 14 of Algorithm \ref{A1}). The weight parameter, $w$ ($0 \leq w \leq 1$) is used to adjust the importance given to stability and accuracy requirements. The U-idx is used to decide if incoming packets warrant ExpIDS to learn a new network traffic distribution. If the need to learn a new distribution is identified, ExpIDS updates the AE hyper-parameters to generate a new classifier. This contrasts with the approach in \cite{enidrift} using an ensemble of sub-classifiers (AE or PCA modules) which updates the weights assigned to those sub-classifiers and may add/delete a sub-classifier based on a generation index value.

We denote the current data chunk as $D_c$ and the training data set for the last AE update as $D_t$. A pool of $K$ AEs is generated by repeatedly training the AE with previous hyper-parameters $K$ times on $M$ data points randomly sampled from $D_t$. The predictions of $K'$ AEs randomly selected from the pool of $K$ AEs are then used to calculate the new U-idx. If the new U-idx is smaller than the previous one, the AE hyper-parameters are updated to a new set selected from a hyper-parameter value grid. If not, the U-idx calculation sub-module waits for a data chunk which meets the stability/accuracy requirements. The hyper-parameters which are updated are the learning rate (step size for updating weights during back-propagation) and the hidden ratio (ratio of number of hidden layers to number of visible layers).

\section{Experimental Results}
\label{results}
In this section, we evaluate the malicious traffic detection performance for \textit{ExpIDS} and compare it with state-of-the-art deep learning-based NIDSs such as Kitsune \cite{kitsune} and ENIDrift-AE \cite{enidrift} for \textit{level-1} and \textit{level-2} of network drift as introduced in \cite{enidrift}. At \textit{level-1} which corresponds to light drift, we use a network traffic dataset collected in a single day and consisting of only one attack type. At \textit{level-2} which corresponds to heavier drift, we use a network traffic dataset collected over multiple days and consisting of different attack types. Kitsune was selected because it has been used as a performance benchmark in prominent NIDS papers published as recently as 2024. ENIDrift has been shown to outperform Kitsune and is the only state-of-the-art NIDS to take network drift into consideration. We use the version of ENIDrift which employs an ensemble of autoencoders, ENIDrift-AE, in our analysis.  

We also generate explanations for \textit{ExpIDS} using xAI methods such as TRUSTEE. We wrote Python scripts for generating the explanation using the TRUSTEE's API and the \textit{ExpIDS} source code. Using the explanations thus generated, we present the top features used by \textit{ExpIDS} to make decisions on input data samples. Finally, we compare the explanations generated for \textit{ExpIDS}, Kitsune and ENIDrift-AE NIDSs by TRUSTEE using the fidelity metric.

There have been newer NIDSs in literature such as HorusEye \cite{horuseye} and IoTA \cite{iot-a} whose detection performance has been benchmarked against Kitsune. However, HorusEye is based on a modified asymmetric autoencoder which has been shown to lack explainability in our previous work \cite{ayush-iot-nids-explain} and does not address concept drift. The source code for IoTA implementation has not been released publicly. Therefore, we have not included HorusEeye and IoTA in our analysis.

\textbf{Datasets}: We use the raw packet captures from Kitsune attack dataset \cite{kitsune} and the CICIDS-2017 dataset \cite{cicids-2017} for training/testing the NIDSs under consideration. The Kitsune dataset consists of significant drift in terms of the different attacks (ARP MiTM, SSDP Flooding, Mirai botnet, etc.) with each attack conducted over a few hours in a day and an imbalance in the ratio of normal and attack traffic. The CICIDS-2017 dataset was collected over five days (July 3-7, 2017) from a small-scale enterprise network. The first day consisted of normal background traffic while the other days consisted of normal and attack traffic, with attacks such as DoS, infiltration, Heartbleed, brute force SSH launched in the network. Though newer IDS datasets have been released such as CICIoV2024 \cite{ciciov2024}, CICEV2023 \cite{cicev2023} and ToN\_IoT \cite{ton-iot}, either they are targeted at specialized applications such as Internet of Vehicles and Electric Vehicle charging infrastructure or they use conventional IT network attack techniques (such as those used in CICIDS-2017) in a different environment (Internet-of-Things).

\textbf{Performance Metrics}: Detection performance of a ML model on a given dataset is measured in terms of precision (PR), recall (RC) and F1 scores. We assume a binary classification problem with positive and negative classes. \textit{Precision} can be expressed as $\frac{TP}{TP+FP}$, where $TP$ is the number of true positives and $FP$ is the number of false positives. It reflects a classifier's capacity to correctly identify true positives without mistakenly labelling negative instances as positive. \textit{Recall} can be expressed as $\frac{TP}{TP+FN}$, where $TP$ is the number of true positives and $FN$ is the number of false negatives. It reflects a classifier's capacity to correctly identify positive instances without mistakenly categorizing them as negative. The \textit{F1 score} is expressed as $2\times\frac{{precision}\times{recall}}{precision+recall}$. It is particularly useful when one needs a balance between precision and recall, rather than focusing solely on one of them.

\textbf{Test Environment}: We used a VMWare ESXi server VM configured with an Intel Xeon Silver 4216 processor running at 2.10GHz, 16GB RAM, 64-bit architecture, 8 cores, and running Ubuntu 18.04/Ubuntu 20.04 OS.

\subsection{Detection Performance}
\textbf{Level 1 drift}: We sample the CICIDS2017-Wed dataset for network packets collected between 10:30 to 11 A.M and use it for our evaluation. The sampled dataset consists of benign traffic interspersed with a a DoS-Hulk attack traffic in the latter half of the collection period. ExpIDS exhibits $4.67\%$ lower precision than ENIDrift-AE which has the highest precision score. However, ExpIDS reports higher recall and F1 scores than ENIDrift-AE and Kitsune NIDSs (See Table \ref{det-perf-table}).

\textbf{Level 2 drift}: For the Kitsune dataset with Mirai botnet traffic, ExpIDS exhibits higher precision, recall and F1 scores than ENIDrift-AE NIDS (See Table \ref{det-perf-table}). However, with $12.4\%$ lower recall and $0.33\%$ lower F1 scores, ExpIDS performs slightly worse than Kitsune NIDS. For the Kitsune datasets with Fuzzing, SSDP flooding, ARP MiTM and SYN DoS attacks, ExpIDS exhibits $25.75\%$, $16.5\%$, $14.42\%$ and $79.2\%$ lower precision respectively than Kitsune NIDS but it shows higher precision than ENIDrift-AE NIDS. In terms of recall, ExpIDS performs equally or better than ENIDrift-AE while both NIDSs perform better than Kitsune. In terms of F1, ExpIDS performs better than ENIDrift-AE and Kitsune NIDSs. Thus, \textit{ExpIDS's attack traffic detection performance is comparable and sometimes better than state-of-the-art NIDSs for both light and significant network drift.}

\begin{table*}
	\centering
	\vspace*{0.1in}
    \begin{tabular}{ | c | c | c | c | c | c | c | c | c | c | c | }
    \hline
    \multicolumn{2}{|c|}{\textbf{Dataset}} & \multicolumn{3}{|c|}{\textbf{ExpIDS}} & \multicolumn{3}{|c|}{\textbf{ENIDrift-AE}} & \multicolumn{3}{|c|}{\textbf{Kitsune}} \\ \cline{3-11}
	\multicolumn{2}{|c|}{} & Precision & Recall & F1 & Precision & Recall & F1 & Precision & Recall & F1 \\ \hline
    \thead{{CICIDS-}\\ {2017}\\ {(Level-1)}} & Wed. & 0.876 (\textcolor{red}{$\downarrow 4.67\%$}) & \textbf{0.966} & \textbf{0.919} & \textbf{0.919} & 0.914 & 0.916 & 0.881 & 0.898 & 0.853 \\ \hline
    \multirow{6}{*}{\thead{{Kitsune}\\ {(Level-2)}}} & Mirai & \textbf{0.940} & 0.876 (\textcolor{red}{$\downarrow 12.4\%$}) & 0.907 (\textcolor{red}{$\downarrow 0.33\%$}) & 0.863 & 0.817 & 0.839 & 0.836 & \textbf{1.000} & \textbf{0.910} \\ \cline{2-11}
    & Fuzz & 0.741 (\textcolor{red}{$\downarrow 25.75\%$}) & \textbf{1.000} & \textbf{0.910} & 0.418 & \textbf{1.000} & 0.590 & \textbf{0.998} & 0.411 & 0.582 \\ \cline{2-11}
    & SSDP & 0.835 (\textcolor{red}{$\downarrow 16.5\%$}) & \textbf{1.000} &\textbf{ 0.852} & 0.611 & \textbf{1.000} & 0.759 & \textbf{1.000} & 0.378 & 0.548 \\ \cline{2-11}
    & ARP & 0.854 (\textcolor{red}{$\downarrow 14.42\%$}) & \textbf{1.000} & \textbf{0.921} & 0.762 & 0.975 & 0.855 & \textbf{0.998} & 0.380 & 0.555 \\ \cline{2-11}
    & SYN & 0.208 (\textcolor{red}{$\downarrow 79.2\%$}) & \textbf{1.000} & \textbf{0.345} & 0.098 & 0.630 & 0.169 & \textbf{1.000} & 0.070 & 0.131 \\ \hline
    \end{tabular}
    \caption{Comparison of detection performance results across \textit{ExpIDS} and state-of-the-art NIDSs}
    \label{det-perf-table}
\end{table*}

\textit{Runtime Performance}: To investigate the runtime performance, we ran ExpIDS with 100 statistical features on a single logical core of an Ubuntu VM running on VMWare ESXi server (details available under test environment in the beginning of this section). On an average, ExpIDS shows a processing speed (training, feature extraction and prediction) of 29.87 packets per second. In comparison, Kitsune can process 300.23 packets per second and ENIDrift-AE can process 36.39 packets per second. ExpIDS's packet processing speed is $17.91\%$ lower than ENIDrift-AE and ENIDrift-AE's packet processing speed is $87.88\%$ lower than Kitsune. ExpIDS's packet processing speed can be increased further by using parallelization techniques and multiple processor cores.

\subsection{TRUSTEE Analysis}
\label{expids-case}

\begin{figure}[h]
\centering
\includegraphics[scale=0.3]{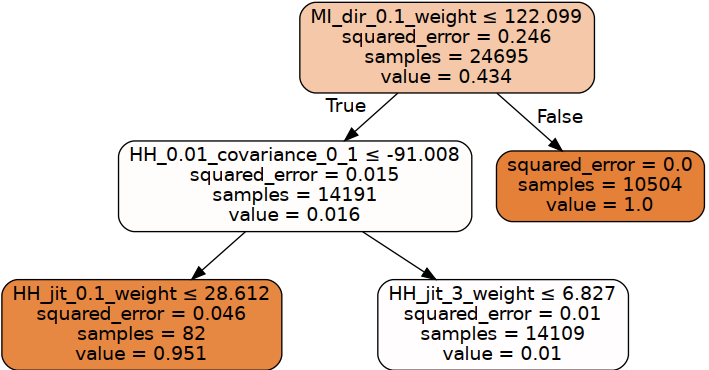}
\caption{Decision tree for \textit{ExpIDS} Mirai model with Top-k pruning (k=10). Only the top 3 layers are shown.}
\label{expids-mirai-dt}
\end{figure}

\textbf{TRUSTEE Explanation}: The results of TRUSTEE analysis for \textit{ExpIDS} are shown in Table \ref{expids-mirai-trustee-table}. Here, \textit{sample size} denotes the fraction of the training dataset which is used to train the student DT model. The DT explanation obtained by applying TRUSTEE with $30\%$ of the original Mirai dataset and skipping pruning achieves a fidelity value of 0.9909 with respected to \textit{ExpIDS}. The DT explanation obtained by applying TRUSTEE with $30\%$ of the original Mirai dataset and top-k pruning ($k = 10$) achieves a fidelity value of 0.964 with respected to \textit{ExpIDS} and is shown in Fig. \ref{expids-mirai-dt}. It is noticeable that pruning a DT's branches leads to a decrease in fidelity value. This is because the resulting smaller trees may overlook essential decision paths.
 
\textit{ExpIDS} uses the following top three prominent features to detect an anomaly: 
\begin{itemize}
    \item $HH\_0.01\_covariance\_0\_1$- Covariance between two packet size streams aggregated by traffic sent between a set of source and destination IP addresses with time window 0.01 (corresponds to 1 minute)
    \item $HH\_0.01\_pcc\_0\_1$- Correlation coefficient between two packet size streams aggregated by traffic sent between a set of source and destination IP addresses with time window 0.01 (corresponds to 1 minute)
    \item $HH\_0.1\_covariance\_0\_1$- Covariance between two packet size streams aggregated by traffic sent between a set of source and destination IP addresses with time window 0.1 (corresponds to 10 seconds) 
\end{itemize}
\textit{Thus, \textit{ExpIDS}'s DT relies mainly on the sizes of packets exchanged per time frame to determine if an attack is underway.}

\begin{table}[h]
	\centering
    \begin{tabular}{ | l | l | l | l | }
    \hline
    \textbf{Sample size} & \thead{\textbf{Top-k pruning}\\ \textbf{used?}} & \textbf{DT size, depth, leaves} & \textbf{Fidelity} \\ \hline
    30\% & No & 85, 26, 43 & 0.9909 \\ \hline
    30\% & Yes (k=10) & 3, 1, 2 & 0.964 \\ \hline
    \end{tabular}
    \caption{TRUSTEE analysis results for \textit{ExpIDS} Mirai model}
    \label{expids-mirai-trustee-table}
\end{table}

\subsection{TRUSTEE Fidelity Comparison}
\label{comp-analysis}
In this sub-section, we use \textit{fidelity} to measure the usefulness of the explanations generated for ExpIDS, ENIDrift-AE and Kitsune by TRUSTEE and compare them. We use the Kitsune Mirai dataset, and compare the fidelity values obtained for our proposed NIDS, \textit{ExpIDS} and the state-of-the-art NIDSs (ENIDrift-AE, Kitsune). With no pruning, ExpIDS achieves the highest fidelity values (0.9909) out of the three NIDSs. With top-k pruning ($k=10$) as well, ExpIDS achieves the highest fidelity values are obtained (0.964) out of the three NIDSs. Thus, \textit{ExpIDS achieves the highest TRUSTEE DT fidelity scores compared with state-of-the-art NIDSs} as pruning is essential to simplify DT explanations, making them more accessible and usable for security practitioners. The fidelity values obtained for the three NIDSs during TRUSTEE analysis have been compared in Table \ref{fidelity-comp-table}.


 
\begin{table}[h]
	\centering
    \begin{tabular}{ | l | l | l | }
    \hline
    \textbf{NIDS} & \thead{\textbf{Fidelity}\\ \textbf{(Without pruning)}} & \thead{\textbf{Fidelity (with top-k}\\ \textbf{pruning, k=10)}} \\ \hline
    ExpIDS & 0.9909 & 0.964 \\ \hline
    ENIDrift-AE & 0.966 & 0.709 \\ \hline
    Kitsune & 0.786 & 0.721 \\ \hline
    \end{tabular}
    \caption{Comparison of TRUSTEE fidelity values with and without pruning for ExpIDS and state-of-the-art NIDSs trained with Kitsune Mirai dataset}
    \label{fidelity-comp-table}
\end{table}



\section{Discussion and Future Work}

The experimental results obtained in the previous section show that \textit{ExpIDS} achieves higher TRUSTEE DT fidelity values than the state-of-the-art NIDSs under consideration, Kitsune and ENIDrift-AE. This was our goal while designing ExpIDS. It reinforces our hypothesis that a simpler NIDS architecture consisting of a single unmodified AE should result in higher DT explanation fidelity values compared to an architecture with modified AE or  an ensemble of AEs. Both Kitsune and ENIDrift-AE use an ensemble of AEs. 
Since AEs are non-linear in nature, TRUSTEE might not be able to build a DT which can imitate the decisions of the black-box AE ensemble within an acceptable error rate. In \cite{ayush-iot-nids-explain}, we had also predicted that there might be a trade-off between ``explainability'' and model performance (e.g., false-positive rate) and NIDS designers need to achieve a balance between the two. As we can observe from the detection performance results comparison between ExpIDS, Kitsune and ENIDrift-AE in Table \ref{det-perf-table}, ExpIDS performs slightly worse in terms of precision for the CICIDS-2017 datasets and for 4 out of 5 of the Kitsune datasets though it outperforms other NIDSs in terms of recall and F1 values. This observation supports our prediction by showing that in turn for better DT explanation fidelity, ExpIDS has to sacrifice marginally in terms of malicious traffic detection precision.

We have also seen from the runtime performance results that \textit{ExpIDS} has a lower packet processing speed than ENIDrift-AE while ENIDrift-AE has a significantly lower packet processing speed compared to Kitsune. This can be attributed to the fact that ExpIDS uses a single de-noising AE whereas both ENIDrift-AE and Kitsune use an ensemble of AEs which helps increase the packet processing rate. Between ENIDrift-AE and Kitsune, ENIDrift-AE takes more time for ground truth label release for newly arrived data chunks and updating the ensemble to address drift. 


In the future, we can deploy ExpIDS on the routers in a small campus/enterprise network and evaluate its real-time  performance (packet processing speed, execution time, memory requirements, detection performance metrics) in the presence of real-world/simulated cyber attacks and real-world network drift. We can also conduct a survey amongst security practitioners for a qualitative study of how useful and trustworthy they find the DT explanations of ExpIDS and if those explanations align with their domain knowledge. 

\section{Related Work}
\label{literature}
The use of explanation techniques on ML systems designed for cybersecurity has been advocated by numerous studies. For instance, the work presented in \cite{ml-insec} has highlighted the significance of explaining the results generated by ML-based detectors deployed in cybersecurity settings. When security practitioners comprehend these explanations, they can improve the security posture of their organization's monitored assets in future operations. While the importance of explainability may be different across sectors, the authors have argued that it is essential for effective application of mitigating actions which address security risks. Similarly, the authors of \cite{dos-n-donts-ml} have suggested using explanation techniques to gain a better understanding of a learning-based security system's features. Although these techniques have their limitations, they can still reveal misleading correlations and allow practitioners to assess their effect on the system's performance. In a survey conducted by the authors of \cite{pragmatic-ml} among security practitioners, participants have indicated that providers of security services should preferably deploy systems that provide easily comprehensible explanations. 

As a result, a few works have proposed ML-based NIDSs and applied explanation techniques on them. 
An NIDS that integrates a CNN-based meta-learner with ensemble learning has been proposed in \cite{ensemble-explain}. This system operates on graphical representations of traffic flows and delivers the necessary level of explainability for its decisions. Additionally, visual depictions of network anomalies have been introduced, enabling practitioners gain valuable insights.
In \cite{spip-ids}, Keshk et al. have proposed an explainable intrusion detection framework for IoT networks employing an LSTM model which is trained and evaluated using features extracted by a combination of explanation techniques.

In \cite{fediot}, the authors have proposed FedIoT, framework that leverages xAI techniques and blockchain to secure Federated Learning (FL)-based IDS in IoT networks. FedIoT uses xAI techniques to identify local model manipulations and mitigate FL-based attacks. It also employs a blockchain-based approach that uses an efficient reputation scheme that ensures the trustworthiness and reliability of the FL training process. Shtayat et al. \cite{iiot-ids} have proposed a deep learning-based NIDS for IIoT networks which combines an extreme-learning machine model and an ensemble of three CNN models. SHAP and LIME techniques have been used to explain the ensemble detector’s decision-making process. In \cite{cps-ids}, the authors have proposed an attack intelligence framework for identifying cyber-physical attacks against physical processes in IIoT environments (e.g., water treatment, gas pipeline) and extracting attack intelligence using various ML/DL algorithms. They also utilize xAI techniques to improve the explainability of the attack attribution module. 

Our work differs from above NIDSs with xAI techniques applied on them \cite{ensemble-explain, spip-ids, fediot, iiot-ids, cps-ids} in the following aspects:
\begin{itemize}
    \item  Those NIDSs do not use the latest xAI tools such as TRUSTEE \cite{trustee} which provide global-level explanations. Instead, they use conventional xAI tools such as SHAP/LIME \cite{shap,lime} which provide local-level explanations by analyzing only a subset of the model’s decisions which might be misleading.
    \item Those NIDSs are not designed to be adaptable to real-world network drift.
\end{itemize}

\section{Conclusion} 
We have proposed a deep learning-based network intrusion detection system, \textit{ExpIDS} designed to have high decision tree explanation fidelity. \textit{ExpIDS} can also adapt to changes in network traffic distribution (drift). Through extensive experiments, we verify that \textit{ExpIDS} achieves a higher TRUSTEE DT fidelity score (0.964) with top-k pruning where $k=10$ while maintaining a malicious traffic detection performance comparable to state-of-the-art NIDSs such ENIDrift-AE and Kitsune for common attacks with light and significant levels of real-world drift.


\bibliographystyle{ieeetran}
\begingroup
\raggedright
\bibliography{xaibib}
\endgroup

\end{document}